\documentclass{appolb}
\usepackage{epsfig}
\usepackage{graphicx}

\begin{document}
\date{\today}
\pagestyle{plain}
\newcount\eLiNe\eLiNe=\inputlineno\advance\eLiNe by -1
\title{{Correlated L\'evy noise in linear dynamical systems} }
\author{Tomasz Srokowski
\address{Institute of Nuclear Physics, Polish Academy of Sciences, PL -- 31-342 Krak\'ow,
Poland }}

\maketitle

\begin{abstract}
Linear dynamical systems, driven by a non-white noise which has the L\'evy 
distribution, are analysed. Noise is modelled by a specific stochastic 
process which is defined by the Langevin equation with a linear force and 
the L\'evy distributed symmetric white noise. Correlation properties of the 
process are discussed. The Fokker-Planck equation driven by that noise is 
solved. Distributions have the L\'evy shape and their width, for a given time, 
is smaller than for processes in the white noise limit. Applicability of 
the adiabatic approximation in the case of the linear force is discussed. 
\end{abstract}
\bigskip

02.50.Ey,05.40.Ca,05.40.Fb

\maketitle

\section{Introduction}

Stochastic dynamical equation (the Langevin equation) describes 
motion of a particle which is subjected to both 
conservative and stochastic force. The latter one can be understood 
either as a result of elimination of internal degrees of freedom or 
as some external physical process. The external noise possesses 
its own time scale and relaxation properties. If relaxation time 
of processes in the environment is relatively short, the white noise 
may be a good approximation: 
the noise variables change rapidly, compared to the particle 
variables. Otherwise the Langevin description must 
involve the correlated ('coloured') noise. 
This problem was widely discussed for the Gaussially distributed 
noise. Well-known physical examples involve 
a phenomenon of narrowing of magnetic resonance lines due to the 
thermal fluctuations \cite{kub} and the fluctuations of dye laser 
light \cite{sho}. The problem of correlated noise also emerges 
when one eliminates some variables in a multi-dimensional dynamical 
system; then the effective low-dimensional description involves 
correlations even if the original many-dimensional system is 
Markovian \cite{hae}. The Langevin equation with the correlated Gaussian 
noise, both additive and multiplicative, is non-Markovian and it 
resolves itself to an integro-differential Fokker-Planck equation which 
can be solved exactly for simple potentials; otherwise approximate 
methods may be applied \cite{hae,gar}. 

Recently, the L\'evy processes -- which constitute a general class 
of the stable processes with the Gaussian process as a special case -- 
attract a considerable interest. They are characterised by long tails, 
which make the variance divergent, and can be observed in many  
systems from various fields: porous and disordered materials, hydrology, biology, 
sociology and finance. Realistic problems are usually characterised by high 
complexity and they exhibit collective phenomena; they involve long-range 
correlations, non-local interactions and a complicated, nonhomogeneous 
(in particular fractal or multifractal) structure of the medium. 
As a result, long jumps may appear and the standard central limit theorem 
is no longer valid. 

It is natural to expect that processes which are driven by a noise with 
long jumps are correlated. As an example can serve an experimental 
study on spontaneous electrical activity of neuronal networks with 
different sizes \cite{seg}. It was found that all networks
exhibited scale-invariant L\'evy distributions. The authors conclude
that different-size networks self-organise to adjust their activities 
over many time scales. The power spectrum, calculated from the experimental 
time series, indicates correlations: it obeys a power-law decay at 
low frequencies for all network sizes.  

The non-Markovian master equation governs probability distributions 
in the framework of the continuous time random walk theory \cite{met1}. 
If jumps are L\'evy distributed, the Fokker-Planck equation is 
fractional both in time and position. 
The integral operators introduce a competition between subdiffusion 
and accelerated diffusion; the latter one results from the infinite 
variance. Integral Fokker-Planck equations were solved for both 
fast and slowly decaying memory kernels \cite{sok}. They can be 
generalised to the fractional orders and to the case of a variable 
diffusion coefficient \cite{sro}. 

In this paper we consider a linear dynamical system which is defined by 
the Langevin equation with the L\'evy distributed non-white noise. 
That problem was solved by H\"anggi and Jung (\cite{hae} and references 
therein) for an arbitrary autocorrelation function in the case of the 
Gaussian noise. However, that approach a priori assumes the 
autocorrelation function and that does not exist if 
$\alpha<2$; we will discuss that difficulty in Sec.II. Therefore we 
introduce a specific model of the correlated noise; we require that 
the model process  should have the L\'evy distribution and be 
correlated (in a sense which will be explained in Sec.II). 
Moreover, it should be as simple as possible.  
We define that process in Sec.II by an adjoint Langevin equation which corresponds 
to the Ornstein-Uhlenbeck process with the white symmetric L\'evy noise. 
We also discuss its correlation properties. The Langevin equation, 
driven by that process, is analysed in Sec.III for simple forms of 
the potential: the free L\'evy motion, the constant force and 
the linear force. Results are summarised in Sec.IV.

\section{Ornstein-Uhlenbeck process with L\'evy noise}

Motion of a particle, which is subjected to the quadratic potential and 
the L\'evy noise, is described by the following linear Langevin equation 
\begin{equation}
\label{la}
\dot\xi(t)=-\gamma\xi(t)+\dot L(t), 
\end{equation}
where the uncorrelated and symmetric noise $L(t)$ is the $\alpha-$stable 
L\'evy process and $\gamma=\mbox{const}>0$. Eq.(\ref{la}), with 
the initial condition $\xi(0)=0$, can be formally solved, 
\begin{equation}
\label{ksiodt}
\xi(t)=\int_0^t K(t-\tau)L(d\tau), 
\end{equation}
where $K(t)=\exp(-\gamma t)$. 
The well-known theory of the Brownian motion corresponds 
to the case $\alpha=2$. Generalisation to the non-Gaussian stable cases, 
which are defined by Eq.(\ref{la}), 
constitutes the Ornstein-Uhlenbeck-L\'evy process (OULP). If $\alpha=2$, 
trajectories are continuous and Eq.(\ref{la}) 
corresponds to the standard Fokker-Planck equation. Otherwise jumps emerge 
and their presence requires introducing integral operators. The Fokker-Planck 
equation, which is suited for problems with jumps, contains the fractional 
operator: 
\begin{equation}
\label{fracou}
\frac{\partial}{\partial t}p(\xi,t)=\gamma\frac{\partial}{\partial\xi}
[\xi p(\xi,t)]+D\frac{\partial^\alpha}{\partial |\xi|^\alpha}p(\xi,t), 
\end{equation}
where $0<\alpha\le2$ denotes the order parameter of the L\'evy distribution 
and $D\ge 0$ is a constant noise intensity. The L\'evy distribution 
itself is given by the following Fourier transform:
\begin{equation}
\label{lev}
P(L)=\frac{1}{\pi} \int_0^\infty \exp(-D k^\alpha)\cos(kL)dk. 
\end{equation}
The density distribution $p(\xi,t)$ can be evaluated either directly from 
Eq.(\ref{ksiodt}) \cite{yan} or by solving Eq.(\ref{fracou}) \cite{jes}. 
The characteristic function reads 
\begin{equation}
\label{solou}
{\widetilde p}(k,t)\equiv\frac{1}{2\pi}\int_{-\infty}^\infty p(\xi,t){\mbox e}^{-ik\xi}d\xi
=\exp\left[-\frac{D}{\alpha\gamma}|k|^\alpha(1-{\mbox e}^{-\gamma\alpha t})\right]. 
\end{equation}
Expression (\ref{solou}) corresponds to the L\'evy stable and 
symmetric process and the width converges with time to a constant, 
producing a stationary distribution. The second moment is divergent, 
unless $\alpha=2$, and also the mean is divergent if $\alpha<1$. 

The Langevin equation driven by the white non-Gaussian noise was studied by 
several authors, both for linear and nonlinear systems 
\cite{jes,garb,dub,vla}. It was generalised to the asymmetric L\'evy 
noise \cite{yan} and to the multiplicative noise \cite{sche,sro1}. 
OULP was also discussed in Ref. \cite{mag} where several fractional 
generalisations were presented.  

Dynamical relation (\ref{la}) introduces a dependence among process 
values $\xi$ at different times: the process 
$\xi(t)$ possesses memory. For the Gaussian case, the autocorrelation 
function serves as a measure of the memory loss. It is defined \cite{gar} 
as the average along a stochastic trajectory: 
\begin{equation}
\label{acor}
G(\tau)=\lim_{T\to\infty}\frac{1}{T}\int_0^T\xi(t)\xi(t+\tau)dt. 
\end{equation}
$G(\tau)$ can be evaluated as the inverse Fourier transform from the spectral function 
\begin{equation}
\label{som}
S(\omega)=\lim_{T\to\infty}\frac{1}{2\pi T}|{\widetilde \xi}(\omega)|^2, 
\end{equation}
where ${\widetilde \xi}(\omega)$ stands for the Fourier transform from $\xi(t)$, 
by means of the Wiener-Khinchin theorem, 
\begin{equation}
\label{wchin}
G(\tau)={\cal F}^{-1}[S(\omega)]. 
\end{equation}
For the ordinary Ornstein-Uhlenbeck process, which is given by 
Eq.(\ref{la}) with $\alpha=2$, the stationary autocovariance function 
$G(\tau)$ follows directly from Eq.(\ref{ksiodt}). It assumes the exponential 
form \cite{gar}, 
\begin{equation}
\label{corou}
G(\tau)=\frac{D}{\gamma}{\mbox e}^{-\gamma|\tau|}, 
\end{equation}
which corresponds to the Lorentzian shape of $S(\omega)$. The correlation 
time $\tau_c=1/\gamma$ measures the decay rate of $G(\tau)$. 

Applying the above formalism to the case $\alpha<2$ is problematic 
since the variance $\sigma^2=G(0)$ becomes infinite. To overcome that difficulty, 
some modifications of the standard covariance definition were introduced. 
One can define \cite{samr,emb} the 'codifference' $\tau_{X,Y}=\sigma_X^\alpha+
\sigma_Y^\alpha-\sigma_{X-Y}^\alpha$, where $X,Y$ are stable and symmetric 
processes. For independent $X$ and $Y$, $\tau_{X,Y}=0$; codifference resolves 
itself to the standard covariance if $\alpha=2$. On the other hand, one can 
utilise the Poissonian structure of the L\'evy process to introduce an infinite 
cascade of Poissonian correlation functions which correspond to the 
autocorrelation function \cite{kla,kla1}. That function depends exponentially 
on time for OULP, Eq.(\ref{la}). Standard correlation formalism of the general L\'evy 
case may be applied if L\'evy measure in the L\'evy-Khinchine formula \cite{pro} 
possesses a cut-off \cite{kla3}; all moments are then finite. Solutions of 
the Langevin equation, which is driven by noise with such a truncated distribution, 
are identical with those for the stable noise up to arbitrarily large distances \cite{tou}. 

The usual definition of the autocorrelation function, Eq.(\ref{acor}), may still be 
applicable to the general stable L\'evy case, despite divergent variance. 
The characteristic function of the increment $\xi(t_2)-\xi(t_1)$ can be 
formally derived \cite{jes}; that function contains all information 
about two-point correlations. 
Special methods of spectral analysis were developed to handle experimental time 
series which involve long jumps, e.g. calculating the count-based periodogram 
\cite{low}. That method allows one to calculate the autocorrelation function 
and power spectrum for long signals, also containing nonstationary trends 
\cite{seg}. We will demonstrate, by means of numerical simulation 
of stochastic trajectories, that speed of memory loss for the process 
(\ref{la}) can be determined by means of the ordinary spectral analysis. 
Let us calculate the power spectrum, Eq.(\ref{som}), from a trajectory which 
follows from Eq.(\ref{la}) and has a given length $T$; the Fourier transform 
is simultaneously evaluated. The relative normalisation of 
$S(\omega)$, $S_0=S(0)\gamma^2$, is finite in any 
calculation since $T$ is always finite. However, it 
depends on $T$ and then cannot be determined, as expected. 
The analysis shows that the quantity $S(\omega)/S_0$ is well determined 
in the limit $T\to\infty$, it obeys the Lorentz function 
\begin{equation}
\label{lor}
\lim_{T\to\infty}S(\omega)/S_0=1/(\gamma^2+\omega^2). 
\end{equation}
The renormalised $S(\omega)$ is presented in Fig.1 for $T=10^4$ 
and some values of $\alpha$ and $\gamma$. All curves follow the Lorentzian 
shape. The value of $S_0$, which emerges from that calculation, may 
be large, it ranges from 1 ($\alpha=2$) to $10^3$ ($\alpha=1.2$). 
\begin{figure}[tbp]
\includegraphics[width=8.5cm]{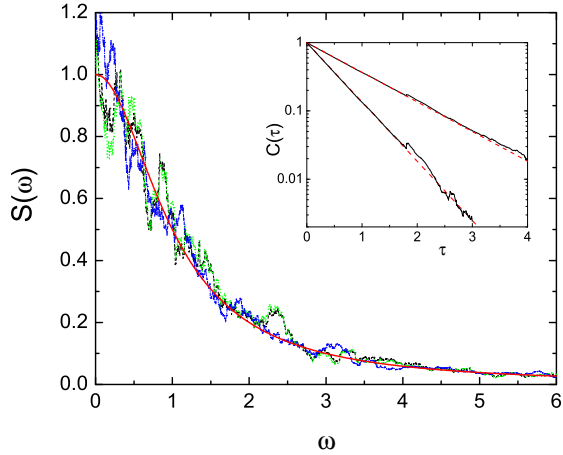}
\caption{Renormalised spectral function for OULP, 
Eq.(\ref{la}), calculated from evolution of a trajectory up to $t=10^4$, 
for the following cases: $\alpha=1.2$ (dashed line), $\alpha=1.5$ 
(green dots) and $\alpha=2$ (blue dashed-dotted line). Red solid line 
denotes the Lorentz function (\ref{lor}). Upper and lower curves 
correspond to $\gamma=1$ and 2, respectively. Inset: ${\cal C}(\tau)$, 
calculated from an ensemble of $10^6$ trajectories with $L_c=10^4$, 
for $\gamma=1$ and 2 (solid lines). Red dashed lines represent the 
function ${\mbox e}^{-\gamma\tau}$.}
\end{figure}

Equivalence of the expression (\ref{acor}) with the ensemble averaged 
covariance is not obvious since a system with long jumps 
may be non-ergodic \cite{reb}. The latter quantity can be directly 
evaluated if one introduces a cut-off in the distribution (\ref{lev}). 
We define the ensemble-averaged autocorrelation function
\begin{equation}
\label{dcov}
{\cal C}(\tau)=\langle \xi(0)\xi(\tau)\rangle/\langle \xi(0)^2\rangle
 \end{equation}
on the assumption that $P(L)=0$ for $L>L_c$. Fig.1 presents that quantity; 
it was derived from the time evolution of individual trajectories by averaging 
over the ensemble. The figure demonstrates that also 
${\cal C}(\tau)$ obeys the exponential dependence (\ref{corou}).

\section{Langevin equation with coloured noise}

In this section we study the stochastic dynamics of a particle which 
is subjected to the L\'evy correlated noise and the linear deterministic 
force. The noise $\xi(t)$ is represented by OULP, 
Eq.(\ref{la}). Then we have to solve a set of two Langevin 
equations,
\begin{eqnarray}
\label{la22}
\dot x(t)&=&f_0-\lambda x(t)+\gamma\xi(t)\nonumber\\
\dot\xi(t)&=&-\gamma\xi(t)+\dot L(t), 
\end{eqnarray}
where $\gamma\ge 0$, $\lambda\ge 0$ and $f_0$ are constants. 
In the presence of jumps, the system remains far from the 
thermal equilibrium and the detailed balance is violated. 
Then $\xi(t)$ can be regarded as an external noise which 
has its own time scale, determined by the parameter 
$\gamma$. In general, processes which obey Langevin equation with 
the correlated noise are non-Markovian since the process values 
are evaluated from mutually dependent noise increments \cite{hae}. 
For large $\gamma$ (short correlation time), $\xi$ is a fast, 
rapidly relaxing variable and the process can be approximated 
by a corresponding white-noise problem, 
by using the methods of adiabatic elimination of fast variables 
\cite{hae,gar}.

\subsection{The case without deterministic force and with a constant force}

The force-free motion, with the white L\'evy noise, is a generalisation 
of the Wiener process; it describes simple diffusion if $\alpha=2$. 
Generalisation to the coloured noise is defined by Eq.(\ref{la22}) 
with $f_0=\lambda=0$. We assume the initial conditions $x(0)=\xi(0)=0$. 
Our aim is to find the probability distribution of the variable $x$. 
One can solve Eq.(\ref{la22}) and utilise the fact that $x(t)$ is still 
a process with independent increments, though multiplied by some 
function of time; then convolution of densities can be performed. 
That method was applied in Ref.\cite{garb} to the second order Langevin 
equation for the case $\alpha=1$. 
We apply van Kampen's method of compound master 
equations \cite{vla} which consists in solving the joint fractional 
Fokker-Planck equation for the two-dimensional system, 
$(x,\xi)$, and integrating over the internal noise $\xi$. That method 
is relatively simple in the case without potential and formally applicable 
also to nonlinear systems with a multiplicative noise. 
In the linear case, the existence, uniqueness and positiveness of 
the solution is ensured \cite{sche}. 

The Langevin equations (\ref{la22}) correspond to the 
fractional Fokker-Planck equation for a joint probability 
distribution $p(x,\xi,t)$ \cite{ris,sche}:
\begin{equation}
\label{frace0}
\frac{\partial}{\partial t}p(x,\xi,t)=\left[-\gamma\frac{\partial}{\partial x}\xi
+\gamma\frac{\partial}{\partial \xi}\xi+D\frac{\partial^\alpha}{\partial |\xi|^\alpha}\right]p(x,\xi,t).
\end{equation}
Knowing the solution of Eq.(\ref{frace0}), the probability distribution 
of the variable $x$ can be obtained by integration over all possible 
realisations of the noise $\xi$: 
\begin{equation}
\label{podx}
p(x,t)=\int_{-\infty}^\infty p(x,\xi,t)d\xi.
\end{equation}
Fourier transformation of Eq.(\ref{frace0}), in respect to both 
$x$ and $\xi$, produces the equation for the characteristic function 
${\widetilde p}(k,\kappa,t)$, 
\begin{equation}
\label{fracek0}
\frac{\partial}{\partial t}{\widetilde p}-\gamma(k-\kappa)\frac{\partial}{\partial\kappa}{\widetilde p}=
-D|\kappa|^\alpha{\widetilde p}, 
\end{equation}
which can be solved exactly by the method of characteristics; 
details are presented in Appendix. The Fourier 
transform of the solution, Eq.(\ref{podx}), follows from Eq.(\ref{A.6}): 
\begin{equation}
\label{sol0}
{\widetilde p}(k,t)={\widetilde p}(k,0,t)={\mbox e}^{-D\sigma(t)|k|^\alpha}, 
\end{equation}
where 
\begin{equation}
\label{s}
\sigma(t)=\frac{1}{\gamma}\int_0^g\frac{\kappa^\alpha}{1-\kappa}d\kappa 
\end{equation}
and $g=1-{\mbox e}^{-\gamma t}$. 
Eq.(\ref{sol0}) predicts the L\'evy shape with the order 
parameter $\alpha$. The width parameter $\sigma(t)$ 
can be estimated in the limit $\gamma t\gg1$, when 
the main contribution to the integral comes from the vicinity 
of the upper integration limit, since then the denominator is 
close to zero: 
\begin{equation}
\label{apro}
\sigma(t)\approx\frac{1}{\gamma}(1-{\mbox e}^{-\gamma t})^\alpha
\int_0^g\frac{d\kappa}{1-\kappa}=t(1-{\mbox e}^{-\gamma t})^\alpha.
\end{equation} 
In the limit $\gamma t\to\infty$, $\sigma$ rises linearly with time and 
$p(x,t)$ coincides with the solution of the uncorrelated problem. 
Convergence to that solution depends on $\alpha$: it is faster for 
smaller $\alpha$. 

The integral (\ref{s}) can be exactly evaluated if $\alpha$ is 
a rational number. In particular, for $\alpha=3/2$ it yields 
\begin{equation}
\label{s32}
\sigma(t)=\frac{2}{\gamma}\left[-(1-{\mbox e}^{-\gamma t})^{1/2}-
(1-{\mbox e}^{-\gamma t})^{3/2}+\mbox{arctanh}\sqrt{1-{\mbox e}^{-\gamma t}}\right]. 
\end{equation}
In the limit $\gamma t\gg1$, the expression (\ref{s32}) predicts a time shift, 
in respect to the white noise case, since it can be approximated by 
$\sigma\approx t-(8/3-2\ln2)/\gamma$. 
\begin{figure}[tbp]
\includegraphics[width=8.5cm]{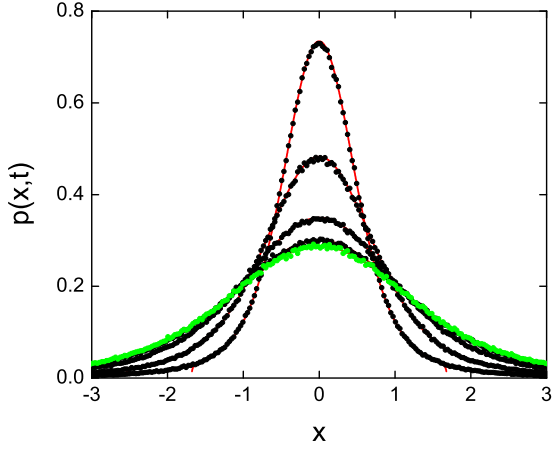}
\caption{Probability distributions at $t=1$ for the force-free 
case calculated by the Monte-Carlo simulations 
(points) for $\gamma=1,2,5,20$ (from top to bottom); the most diffused 
case corresponds to the white noise limit ($\gamma=\infty$). Analytical 
results, calculated from Eq.(\ref{sze}) with $\sigma$ from Eq.(\ref{s32}), 
are presented as solid lines. The order parameter $\alpha=1.5$. Numerical 
simulations were performed with the time step $\tau=0.005$ and averaged 
over $10^7$ events.}
\end{figure}

Numerical values of the probability distribution $p(x,t)$, which result 
from inversion of the characteristic function (\ref{sol0}), 
can be obtained from the series expansion \cite{sch},
\begin{equation}
\label{sze}
p(x,t)=\frac{1}{\pi \sigma^{1/\alpha}\alpha}\sum_{n=0}^\infty
\frac{\Gamma[1+(2n+1)/\alpha]}{(2n+1)!!}(-
1)^n\left(\frac{x}{\sigma^{1/\alpha}}\right)^{2n},
\end{equation}
if $|x|$ is not too large. Fig.2 presents those distributions for 
the case $\alpha=1.5$ at $t=1$, $\sigma(t)$ was calculated from 
Eq.(\ref{s32}). Figure shows that the memory affects the rate of 
spreading of the distribution: $p(x,t)$ is broadest for the white noise 
case, $\gamma=\infty$, and it contracts to the delta function in 
the limit $\gamma\to0$. Results are compared with the Monte Carlo 
simulations of individual trajectories, according to the stochastic  
equations (\ref{la22}). For that purpose, a simple Euler algorithm 
was applied. The white noise value 
at $i-$th integration step, $L_i$, was represented by the term  
$\tau^{1/\alpha}L_i$, where $\tau$ was the step size \cite{jan}. 
Probability distributions were obtained by averaging over an 
statistical ensemble of the individual trajectories. 
Since the analytical result does not contain any approximation, 
agreement with the simulations is exact. 

Problem of the linear potential, $-f_0 x$, where 
$f_0=$const., can be reduced to the force-free case which was 
discussed above. The first equation in Eq.(\ref{la22}) takes the form 
$\dot x(t)=f_0+\gamma\xi(t)$.
From the corresponding fractional Fokker-Planck equation, 
\begin{equation}
\label{frace1}
\frac{\partial}{\partial t}p(x,\xi,t)=\left[-\frac{\partial}{\partial x}(f_0+\gamma\xi)
+\gamma\frac{\partial}{\partial \xi}\xi+D\frac{\partial^\alpha}{\partial |\xi|^\alpha}\right]p(x,\xi,t),
\end{equation}
we derive equation for the characteristic function: 
\begin{equation}
\label{fracek1}
\frac{\partial}{\partial t}{\widetilde p}-\gamma(k-\kappa)\frac{\partial}{\partial\kappa}{\widetilde p}=
-(if_0k+D|\kappa|^\alpha){\widetilde p}. 
\end{equation}  
Its solution, ${\widetilde p}(k,\kappa,t)={\mbox e}^{-if_0kt}{\widetilde p}_0$, 
where ${\widetilde p}_0$ is given by Eq.(\ref{A.6}), follows from the general 
theory \cite{sche}. It can be also obtained by separation of real and imaginary 
parts of ${\widetilde p}(k,\kappa,t)$ and by solving the resulting set of two equations. 
Integration over the variable $\xi$ produces the final result: 
\begin{equation}
\label{sol1}
{\widetilde p}(k,t)={\mbox e}^{-if_0kt}{\widetilde p}_0, 
\end{equation}
where ${\widetilde p}_0(k,t)$ follows from Eq.(\ref{sol0}). The distribution 
$p(x,t)$ has the same shape, for any time, as that for the case $f_0=0$ but 
it is shifted by $f_0t$. That means that the average rises linearly with time, 
$\langle\xi\rangle=f_0t$ (if $\alpha>1$), and the distribution widens 
with time according to the function $\sigma(t)$, Eq.(\ref{s}). In the limit 
$\gamma\to0$, $p_0(x,t)=\delta(x)$ which corresponds to a deterministic 
motion with velocity $f_0$. 
Probability distributions which follow from the Monte Carlo simulations 
(not presented) agree with the solution (\ref{sol1}). 

In the limit $\gamma t\to\infty$, Eq.(\ref{sol1}) 
coincides with the solution of fractional Fokker-Planck equation 
with the constant force for the white noise case \cite{jes}. 
The problem of transport in an effective constant force field emerges 
in the framework of the continuous time random walk theory when one 
considers a biased walk \cite{met}. It resolves itself 
to the fractional Fokker-Planck equation with a drift term.

\subsection{Linear force}

\begin{figure}[tbp]
\includegraphics[width=8.5cm]{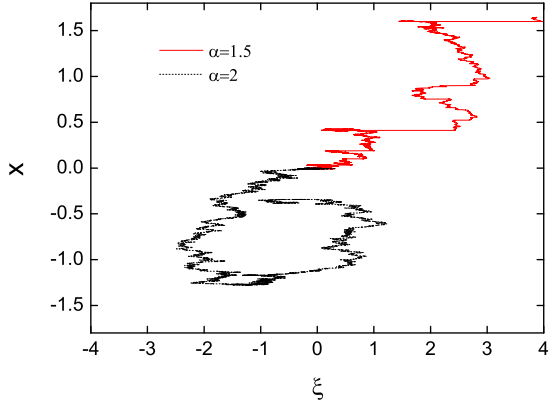}
\caption{Exemplary stochastic trajectories in the space 
$(\xi,x)$, calculated from Eq.(\ref{la22}) with time step 
$\tau=5\cdot10^{-4}$ up to $t=3$, for $\lambda=1$ and $\gamma=1$. 
The trajectory for the case $\alpha=1.5$ is positioned in upper-right 
quarter of the figure.} 
\end{figure}

The system is defined by Eq.(\ref{la22}) with $f_0=0$, 
where $\lambda>0$ measures intensity of the deterministic force. 
The aim of this section is a comparison of exact probability distributions, 
obtained by numerical simulation of two-dimensional stochastic trajectories 
from Eq.(\ref{la22}), with predictions of the adiabatic approximation. 
\begin{figure}[tbp]
\includegraphics[width=8.5cm]{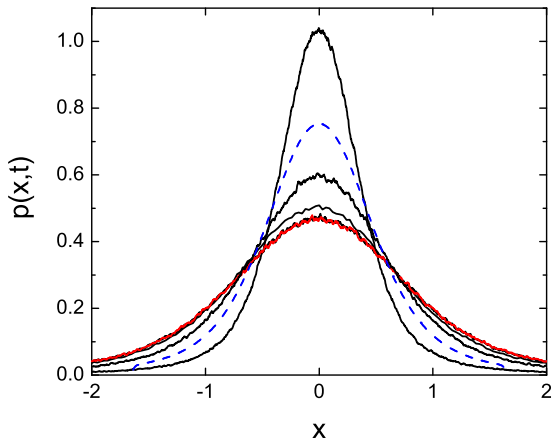}
\caption{Time evolution of the probability 
distribution for the system with linear force, Eq.(\ref{la22}), 
calculated for the following times: 1, 2, 3, 5 (black 
solid lines from top to bottom). The case $t=10$, which corresponds 
to the stationary solution, is marked by red solid line. The stationary 
solution which is predicted by the adiabatic approximation, 
Eq.(\ref{adia2}), is shown as blue dashed line. The other parameters: 
$\alpha=1.5$, $\lambda=1$ and $\gamma=1$.} 
\end{figure}

Fig.3 presents examples of stochastic trajectories for 
two cases: the L\'evy distribution with $\alpha=1.5$ and 
for the normal distribution. In the former case, large jumps, 
typical for the L\'evy processes, are visible along the horizontal 
direction which represents OULP (Eq.(\ref{la})). 
The process $x(t)$, in turn, is stronger localised 
for both values of $\alpha$. The plot shrinks in the horizontal 
direction with increasing $\gamma$ (not shown) which reflects 
the fact that $\xi$ becomes the fast variable: it relaxes rapidly to 
$\xi=0$. Averaging over a large number of trajectories produces the probability 
distribution $p(x,t)$. Fig.4 demonstrates that it converges with time 
to the stationary distribution, as in the white noise case. The 
time which is needed to reach the steady state equals 5 for the case 
presented in the figure. The shape of $p(x,t)$ coincides with the L\'evy 
distribution for any $\gamma$ and its order parameter $\alpha$ corresponds 
to that of the driving noise $L(t)$. The apparent width rises with 
$\gamma$ and, for large $\gamma$, the white-noise limit is reached. 

To estimate the dependence $\sigma(\gamma)$ the characteristic function 
$\exp(-\sigma(t)|k|^\alpha)$ was evaluated. Results 
are presented in Fig.5. The distribution very slowly converges with 
$\gamma$ to the white-noise value whereas it shrinks to the delta function 
for $\gamma\to0$. 
\begin{figure}[tbp]
\includegraphics[width=8.5cm]{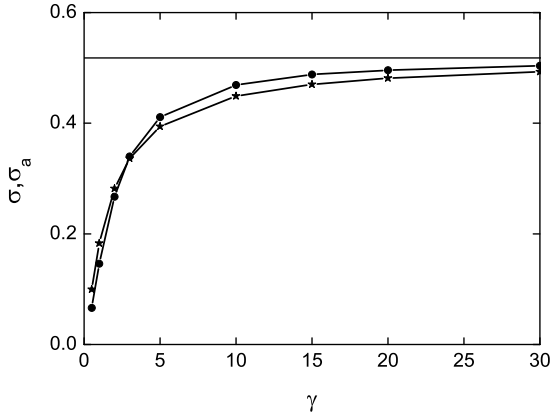}
\caption{Width parameter $\sigma$, evaluated from the 
characteristic function for $t=1$, as a function of memory parameter 
$\gamma$ (points). Results of the adiabatic approximation, Eq.(\ref{siga}), 
are marked by stars. The parameters are: $\alpha=1.5$ 
and $\lambda=1$. Horizontal line marks the white noise limit.}
\end{figure}

The adiabatic approximation in the case of the normally distributed noise 
was discussed in Ref.\cite{jun}; we apply a similar procedure. Combination 
of equations (\ref{la22}) yields a single second order stochastic equation:
\begin{equation}
\label{adia}
\ddot x(t)=-(\lambda+\gamma)\dot x(t)-\lambda\gamma x(t)+\gamma \dot L(t). 
\end{equation}
One can demonstrate, by introducing a new time variable $t'=\sqrt{\gamma}t$, 
that the term $\ddot x$ is small both for $\gamma\to0$ and $\infty$. 
Therefore, Eq.(\ref{adia}) can be approximated by the following equation 
\begin{equation}
\label{adia2}
\dot x(t)=-\lambda c_\gamma x(t)+c_\gamma \dot L(t),   
\end{equation}
where $c_\gamma=1/(1+\lambda/\gamma)$. The corresponding fractional 
Fokker-Planck equation is analogous to Eq.(\ref{fracou}) and it can be 
easily solved. Fourier transform of the solution is 
${\widetilde p}_a(k,t)=\exp(-\sigma_a(t)|k|^\alpha)$, where the apparent width 
\begin{equation}
\label{siga}
\sigma_a(t)=\frac{c_\gamma^\alpha D}{\alpha\lambda} (1-{\mbox e}^{-\alpha\lambda t}). 
\end{equation}
The adiabatic solution, $p_a(x,t)$, converges with time to the steady 
state and it coincides with the uncorrelated process in the limit 
$\gamma\to\infty$; Eq.(\ref{siga}) implies that $\sigma_a$ rises with $\gamma$. 
Eq.(\ref{adia2}) is exact both for $\gamma\to0$ -- when the delta function 
is the solution -- and in the limit $\gamma\to\infty$ (the Smoluchowski limit). 
For intermediate values of $\gamma$, one can expect that Eq.(\ref{adia2}) 
is a good approximation on time scales $t>1/(\lambda+\gamma)$ and at 
distances $\gg D^{-1/2}/(\gamma^{1/2}+\lambda\gamma^{-1/2})$ \cite{hae}. 

The width parameter $\sigma(t)$ for the exact solution is compared with 
$\sigma_a$, predicted by Eq.(\ref{siga}), in Fig.5. Some differences are 
visible but qualitative agreement of the functions $\sigma(\gamma)$ for 
both cases is good in the entire range of presented $\gamma$ values. 
In general, however, 
discrepancies may be more pronounced. For example, the 
adiabatic approximation underestimates the width of the 
steady-state distribution for $\gamma=1$, which is shown 
in Fig.4, by a factor of two (0.24 vs. 0.48).

\section{Summary and conclusions}

We have studied the linear dynamical systems which are driven by  
the additive, non-white L\'evy noise. That noise 
is modelled by a concrete, simple stochastic process, OULP. Then 
the system is defined in terms of two Langevin equations. 
OULP reveals the memory effects, as for the ordinary 
Ornstein-Uhlenbeck process, but their quantitative description is more 
difficult because of the divergent variance. 
We have presented a numerical example which demonstrates that 
the renormalised autocorrelation function $G(t)$ can be useful 
as a measure of the memory loss; 
it falls exponentially with time for any order parameter $\alpha$. 
The same result was obtained for the ensemble-averaged autocorrelation 
function on the assumption that the L\'evy distribution is truncated. 

In the absence of any deterministic force, the non-Markovian problem resolves 
itself to the Wiener-L\'evy process (correlated L\'evy motion). 
The resulting probability distribution has the L\'evy shape, 
with parameter $\alpha$, and it converges with time to that for the uncorrelated 
case. Correlation time $\tau_c=1/\gamma$ determines the distribution width: 
the larger $\tau_c$, the narrower the distribution. The case 
of the constant force $f_0$ is similar; shape and width of the distribution is 
the same but the time-dependent shift $f_0t$ emerges. 

Solution for the case of the linear force converges with time to the steady state, 
as for the white-noise problem, and its shape is L\'evy with parameter 
$\alpha$. Inclusion the finite correlation time narrows the distribution, 
analogously to the case without a force. The above observations agree with 
the adiabatic approximation approach. That method 
deals with a corresponding, effective white-noise process and resolves itself 
to the Langevin equation of the first order. It is supposed to be accurate if 
$\gamma$ is sufficiently large or if $\gamma\to0$. For intermediate values of 
$\gamma$, overall predictions of the adiabatic approximation in 
respect to the distribution shape and its dependence on $\gamma$ 
are still correct, nevertheless some quantitative discrepancies 
have been found.

\section*{APPENDIX}

\setcounter{equation}{0}
\renewcommand{\theequation}{A\arabic{equation}} 

In the Appendix, we solve the fractional Fokker-Planck equation, Eq.(\ref{fracek0}),  
by means of the method of characteristics. 

First, we put the equation into the form 
\begin{equation}
\label{A.1}
|\kappa|^{-\alpha}\frac{\partial}{\partial t}{\widetilde p}(k,\kappa,t)-
\gamma(k-\kappa)|\kappa|^{-\alpha}\frac{\partial}{\partial\kappa}{\widetilde p}(k,\kappa,t)
=-D{\widetilde p}(k,\kappa,t). 
\end{equation}
Eq.(\ref{A.1}) is the linear partial differential equation of 
the first order with only two variables, $t$ and $\kappa$, since $k$ 
can be regarded as a constant parameter. The equation can be handled by 
the method of characteristics \cite{eva}. 
The method consists in reducing the problem to solution of a system 
of ordinary differential equations (characteristic equations). 
Those equations determine variables $t$, $\xi$ and $z$, as functions of parameters 
$s$ and $r$, on a characteristic curve. They are of the form 
\begin{eqnarray}
\label{A.2}
\frac{dt(r,s)}{ds}&=&|\kappa|^{-\alpha}\nonumber\\
\frac{d\kappa(r,s)}{ds}&=&-\gamma(k-\kappa)|\kappa|^{-\alpha}\\
\frac{dz(r,s)}{ds}&=&-Dz\nonumber
\end{eqnarray}
with the initial conditions
\begin{eqnarray}
\label{A.3}
t(r,0)&=&0\nonumber\\
\kappa(r,0)&=&r\\
z(r,0)&=&1;\nonumber
\end{eqnarray}
the third condition reflects the requirement that $p(x,\xi,0)$ is to be 
the delta function in the variable $\xi$. 
We must solve the system (\ref{A.2}) and then eliminate the parameters 
$r(t,\kappa)$ and $s(t,\kappa)$. The final solution of Eq.(\ref{A.1}) 
is given by ${\widetilde p}(k,\kappa,t)=z(r,s)$. 
Combination of the first and second equation gives the relation 
between $t$ and $\kappa$ on the characteristic curve: 
$t=\ln[(\kappa-k)/(r-k)]/\gamma$, where the initial conditions 
(\ref{A.3}) were taken into account. The above relation determines 
the parameter $r$: 
\begin{equation}
\label{A.4}
r(t,\kappa)=k-(k-\kappa){\mbox e}^{-\gamma t}. 
\end{equation}
Integration of the third equation (\ref{A.2}) is straightforward, 
$z(r,s)={\mbox e}^{-Ds}$, and $s$, as a function of the variables 
$\kappa$ and $t$, follows from the second equation: 
\begin{equation}
\label{A.5}
s(t,\kappa)=\frac{1}{\gamma}\int_r^\kappa\frac{|\kappa'|^\alpha}{\kappa'-k}d\kappa'. 
\end{equation}
The final solution reads 
\begin{equation}
\label{A.6}
{\widetilde p}(k,\kappa,t)={\mbox e}^{-Ds}, 
\end{equation}
where $s$ is given by Eq.(\ref{A.5}). The solution (\ref{A.6}) can be 
verified by a direct inserting into Eq.(\ref{A.1}) and applying the Leibniz rule 
for differentiation of the integral.


\begin{thebibliography}{99}

\bibitem{kub}
R. Kubo, {\it Fluctuation, Relaxation and Resonance in Magnetic Systems} 
(Oliver and Boyd, London, 1982). 

\bibitem{sho}
R. Short, L. Mandel, and R. Roy, {\it Phys. Rev. Lett.} {\bf 49}, 647 (1982). 

\bibitem{hae}
P. H\"anggi and P. Jung, Adv. Chem. Phys., Vol. LXXXIX, 
Ed. by I. Prigodine and Stuart A. Rice (John Wiley \& Sons, 1995). 

\bibitem{gar}
C. W. Gardiner, {\it Handbook of Stochastic Methods for Physics, Chemistry
and the Natural Sciences} (Springer-Verlag, Berlin, 1985).

\bibitem{seg}
R. Segev, M. Benveniste, E. Hulata, N. Cohen, A. Palevski, E. Kapon, 
Y. Shapira, and E. Ben-Jacob, {\it Phys. Rev. Lett.} {\bf 88}, 118102 (2002). 

\bibitem{met1}
R. Metzler and J. Klafter, {\it Phys. Rep.} {\bf 339}, 1 (2000). 

\bibitem{sok}
I. M. Sokolov, {\it Phys. Rev.} {\bf E66}, 041101 (2002). 

\bibitem{sro}
T. Srokowski, {\it Phys. Rev.} {\bf E75}, 051105 (2007).

\bibitem{yan}
V. V. Yanovsky, A. V. Chechkin, D. Schertzer, and A. V. Tur,
{\it Physica} {\bf A282}, 13 (2000).

\bibitem{jes}
S. Jespersen, R. Metzler, and H. C. Fogedby, 
{\it Phys. Rev.} {\bf E59}, 2736 (1999). 

\bibitem{garb}
P. Garbaczewski and R. Olkiewicz, {\it J. Math. Phys.} {\bf 41}, 6843 (2000). 

\bibitem{dub}
A. A. Dubkov, B. Spagnolo, and V. V. Uchaikin, 
{\it Intern. Journ. of Bifurcation and Chaos} {\bf 18}, 2649 (2008).

\bibitem{vla}
M. O. Vlad, M. G. Velarde, and J.Ross, {\it J. Math. Phys.} {\bf 45}, 736 (2004).

\bibitem{sche}
D. Schertzer, M. Larchev\^{e}que, J. Duan, V. V. Yanovsky, and S. Lovejoy, 
{\it J. Math. Phys.} {\bf 42}, 200 (2001). 

\bibitem{sro1}
T. Srokowski, {\it Phys. Rev.} {\bf E79}, 140104(R) (2009); 
{\it ibid.} {\bf E80}, 051113 (2009). 

\bibitem{mag}
M. Magdziarz, {\it Physica} {\bf A387}, 123 (2008). 

\bibitem{samr}
G. Samrodintsky and M.S. Taqqu, {\it Stable Non-Gaussian Random Processes} (Chapman \& Hall, London, 1994).

\bibitem{emb}
P. Embrechts and M. Maejima, {\it Selfsimilar Processes} (Princeton University Press, Princeton, 2002). 

\bibitem{kla}
I. Eliazar and J. Klafter, {\it Physica} {\bf A376}, 1 (2007). 

\bibitem{kla1}
I. Eliazar and J. Klafter, {\it J. Phys.} {\bf A40}, F307 (2007). 

\bibitem{pro}
P. E. Protter, {\it Stochastic Integration and Differential Equations} 
(Springer-Verlag, Berlin, 2005). 

\bibitem{kla3}
I. Eliazar and J. Klafter, J. Stat. Phys. {\bf 119}, 165 (2005).

\bibitem{tou}
H. Touchette and E. G. D. Cohen, {\it Phys. Rev.} {\bf E80}, 011114 (2009).

\bibitem{low}
S. B. Lowen, T. Ozaki, E. Kaplan, B. E. A. Saleh, and M. C. Teich, 
{\it Methods} {\bf 24}, 377 (2001).

\bibitem{reb}
A. Rebenshtok and E Barkai, {\it J. Stat. Phys.} {\bf 133}, 565 (2008). 

\bibitem{ris}
H. Risken, {\it The Fokker-Planck Equation} (Springer-Verlag, Berlin,
1996).

\bibitem{sch}
W. R. Schneider, in {\it Stochastic Processes in Classical and Quantum Systems, 
Lecture Notes in Physics}, edited by S. Albeverio, G. Casati, D. Merlini 
(Springer, Berlin, 1986), Vol. 262. 

\bibitem{jan}
A. Janicki and A. Weron, {\it Simulation and Chaotic Behavior of
$\alpha$-Stable Stochastic Processes} (Marcel Dekker, New York,
1994). 

\bibitem{met}
R. Metzler, J. Klafter, and I. M. Sokolov, {\it Phys. Rev.} {\bf E58}, 1621 (1998). 

\bibitem{jun}
P. Jung and P. H\"anggi, {\it Phys. Rev.} {\bf A35}, 4464 (1987). 

\bibitem{eva} 
L. C. Evans, {\it Partial Differential Equations} 
(American Mathematical Society, Providence, Rhode Island, 1998). 

\end{thebibliography}
\end{document}